\documentclass[10pt]{iopart}

\usepackage{amsfonts}
\usepackage{amssymb}
\usepackage{epsfig}
\usepackage{subfigure}
\usepackage{graphicx}
\usepackage{color}

\def\beq{\begin{equation}}
\def\eeq{\end{equation}}
\def\beqa{\begin{eqnarray}}
\def\eeqa{\end{eqnarray}}
\def\hf{\textstyle{1\over2}}
\def\3hf{\textstyle{\frac{3}{2}}}

\newcommand{\ket}[1]{\vert #1 \rangle}

\newcommand{\bra}[1]{\langle #1 \vert}

\renewcommand{\e}{\hbox{\rm e}}
\newcommand{\myfrac}[2]{\leavevmode\kern.1em\raise.5ex\hbox{\scriptsize
$#1$}\kern-.1em {\scriptsize
/}\kern-0.10em\lower.25ex\hbox{\scriptsize $#2$}}

\newcommand{\op}[1]{\hat{#1}}

\begin{document}

\title{Qutrit squeezing via semiclassical evolution}
\author{Andrei B. Klimov$\,^1$, Hossein Tavakoli Dinani$\,^2$, \newline Zachari
E.D. Medendorp$\,^2$ and Hubert de Guise$\,^2$}

\begin{abstract}
We introduce a concept of squeezing in collective qutrit systems through a
geometrical picture connected to the deformation of the isotropic
fluctuations of $su(3)$ operators when evaluated in a coherent state. This
kind of squeezing can be generated by Hamiltonians non-linear in the
generators of $su(3)$ algebra. A simplest model of such non-linear evolution
is analyzed in terms of semiclassical evolution of the $SU(3)$ Wigner
function.
\end{abstract}


\address{$\,^1$ Departamento de F\'{\i}sica, Universidad de
Guadalajara, 44420 Guadalajara, Jalisco, Mexico } 
\address{$\,^2$ Department of Physics, Lakehead University, Thunder
Bay, Ontario P7B 5E1, Canada}


\section{Introduction}

The concept of squeezing in different systems has attracted significant
attention due to its transparent physical meaning, related to the reduction
of quantum fluctuations below some given threshold. Although most of
applications of squeezing are related to the improvement of measurements
precision, squeezing intrinsically reflects the existence of some particular
correlations between parts of a quantum system. Since the squeezing
parameters contains easily measurable first and second order moments of
collective operators, this entails a successful application of squeezing
criteria to detect quantum entanglement \cite{sorensen}, \cite{briegel}, 
\cite{rev}. 

Historically, much attention has been paid to squeezing of the
electromagnetic field modes or squeezing in $SU(2)$ - or spin-like - systems. Recently, more complex
experiments on quantum systems having higher symmetries have been proposed,
particularly in relation to some possible applications to quantum
information processes. Candidate qutrit systems described by the group $%
SU(3) $ include Bose-Einstein condensates and three--level atomic ensembles
interacting with quantized fields.

The definition of squeezing, while universal for harmonic oscillator--like
systems, is otherwise far from unique. In spin-like systems there are
several approaches used to define a squeezing parameters \cite{SS,sorensen,
lewenstein, luis, steinberg, devi, rev}. All parameters compare fluctuations
of some suitably chosen observables with a certain threshold given by
fluctuations in some reference state (or family of states). The coherent
states of the corresponding quantum system are often taken as the family of
reference states.

One of the crucial properties of coherent states is the invariance of the
fluctuations of some observables under certain type of continuous
transformations.  This property allows the definition of the so-called Quantum Standard Limit 
\cite{win}.

In this article we use this property of coherent states to introduce the
concept of squeezing for systems with $SU(3)$ symmetry (the extension to
systems with $SU(N)$ symmetry can also be done). The main idea consists in
defining the full family ${\cal K}$ of collective operators (which in practice are some
linear combinations of generators of the $su(3)$ algebra) for which the
fluctuations evaluated using $SU(3)$ coherent states are invariant under the
same group transformation that leaves invariant the fiducial state used to
construct the set of coherent states.

We will show that, for a Hilbert space carrying an irreducible
representation of $SU(3)$ of the symmetric type, we can use $3$ continuous parameters $\alpha_3,\beta_3,\chi$ to
label a generic element  ${{\cal K}}(\alpha_3,\beta_3,\chi)\in {\cal K}$,  but fluctuations of 
${{\cal K}}(\alpha_3,\beta_3,\chi)$, 
when evaluated using a suitable $SU(3)$ coherent state, are isotropic, \emph{i.e.} do not depend on $\alpha_3,\beta_3,\chi$. 
Considering these (invariant)
fluctuations as defining our threshold, we introduce squeezing as a
reduction of fluctuations below the limit of these isotropic fluctuations in
coherent states.

Since our objective is to show how $SU(3)$ squeezing can emerge rather than propose a general criterion,
we will focus on the deformations of probability distributions resulting from the Hamiltonian evolution 
of an initial coherent state. 
Geometrically, a group transformation obtained by exponentiating a linear combination of generators and acting on a state produces a simple rigid displacement of the associated probability distribution and is not associated with
the introduction of correlations.  A deformation of the probability
density does mean that quantum correlations between parts of the system are
generated; hence quantum correlations which generate the squeezing can only arise from 
non-linear interactions.

As the characteristic
times needed to produce such correlations are inversely proportional to some
power of the dimension of the system, correlations develop very rapidly and
the analysis can be done using semi-classical methods. In this article we
will use the $SU(3)$ Wigner function method \cite{GA} to describe a
non-linear evolution of a quantum system with the $SU(3)$ symmetry group.

The article is organized as follows: in Section II we briefly recall general
ideas on the coherent states for systems with $SU(2)$ and $SU(3)$ symmetries
and construct the operators with isotropic fluctuations in the corresponding
coherent states. In Section III we analyze squeezing generated by a simple
non-linear $SU(3)$ Hamiltonian. In Section IV the $SU(3)$ Wigner function
formalism is presented and applied to find the evolution of the squeezing
parameter under the non-linear Hamiltonian.

\section{Coherent states}

Following the general construction \cite{perelomov, gilmore} a coherent
state for a system with a given symmetry group $\mathcal{G}$ acting irreducibly
in a Hilbert space $\mathbb{H}$ is defined as a fiducial state
displaced by a group transformation in $\mathcal{G}$. We take this fiducial state to be 
the highest weight state of the irreducible representation carried by $\mathbb{H}$. 
The highest weight state is invariant (up to a phase) under transformation from the subgroup $\mathcal{H}\subset 
\mathcal{G}$, so displacements of this state are labelled by points $\Omega$ on the coset
$\mathcal{G}/\mathcal{H}$. The latter is known to be the classical
phase space of the corresponding quantum system \cite{Onofri}.

Below, we briefly review coherent states for the $SU(2)$ and $SU(3)$ groups, focusing only on the
symmetric representations. In this case a coherent state can be
considered as a composite state, occurring as a direct
product of identical "single particle" states of systems with 2 or 3 energy
levels, and invariant under permutation of the ``particle'' labels.  In other words, coherent states can be conveniently thought of as
symmetric (under permutation of particles) factorized states, thus
displaying maximal \emph{classical} correlations. Given any
coherent state we can always find a operator written as linear combination
of generators such that the fluctuations of this operator evaluated in the
coherent state is invariant with respect to the transformations generated by
the stationary subgroup $\mathcal{H}$. Moreover, the fluctuations of this
operator reach a value determined by the dimension of $\mathbb{H}$.

\subsection{$SU(2)$ coherent states}

The $su(2)$ algebra is spanned by $\{\op S_+,\op S_-,\op S_z\}$, with non-zero commutation relation
\beq
[\op S_z,\op S_{\pm}]=\pm \op S_{\pm}\, ,\qquad
[\op S_+,\op S_-]=2\op S_z\, .
\eeq
A basis for the irrep $j$ of dimension $2j+1$ is spanned by the states $\{\ket{jm},m=-j,\ldots,j\}$.  The basis states
satisfy
\beq
\hat S_{\pm}\ket{jm}=\sqrt{(j\mp m)(j\pm m+1)}\ket{j,m\pm 1}\, ,\quad
\op S_z\ket{jm}=m\ket{jm}\, .
\eeq
The highest weight state is $\ket{jj}$.  It is invariant (up to a phase) under the subgroup ${\cal H}=\{T(\gamma)
\equiv\e^{-i\gamma \op S_z}\}$.
The parameter $\gamma$ ranges between $0$ and $2\pi$ when $2j$ is even, and between $0$ and $4\pi$ when $2j$ is odd.
We can now define a family ${\cal S}$ of observables through
\beq
{\cal S}=\{T(\chi)\,\op S_x\,T^{-1}(\chi)\}\, ,\qquad T(\chi)=\e^{-i\chi \op S_z}\in {\cal H}\, .
\eeq
A typical element of the family is
\beq
\op S(\chi)\equiv T(\chi)\,\op S_x\,T^{-1}(\chi)=\op S_x\,\cos\chi+\op S_y\,\sin\chi\, .
\eeq
For any $\op S (\chi)\in\cal{S}$ we find, using $\ket{jj}$, that $(\Delta \op S (\chi))^2=j$, independent of the 
element $T(\chi)\in{\cal H}$.

The standard set $\{|\vartheta ,\varphi \rangle \}$ of $SU(2)$ coherent
states of angular momentum $j$ is defined as%
\begin{equation}
|\vartheta ,\varphi \rangle =D\left( \vartheta ,\varphi \right) |j,j\rangle ,
\label{su2_cs}
\end{equation}%
where $D\left( \vartheta ,\varphi \right) =\exp (-\frac{\vartheta }{2}%
(\e^{-i\varphi /2}\hat{S}_{+}-\e^{i\varphi /2}\hat{S}_{-}))$. 
The coherent states (\ref{su2_cs}) can be represented as a product of $2j$
one-qubit states, 
\beqa
|\vartheta ,\varphi \rangle &\propto& |\vartheta ,\varphi \rangle _{1}\otimes
|\vartheta ,\varphi \rangle _{2}\otimes \,\ldots \,\otimes |\vartheta
,\varphi \rangle _{n}\,, \\
&&|\vartheta ,\varphi \rangle _{a}\equiv \e^{i\varphi /2}\sin \textstyle{\frac{1}{2}}%
\vartheta |+\hf\rangle _{a}+\e^{-i\varphi /2}\cos \textstyle{\frac{1}{2}}%
\vartheta |-\hf\rangle _{a}\, .
\eeqa
$|\vartheta ,\varphi \rangle$ is completely specified geometrically through the direction 
$\vec{n}=(n_{x},n_{y},n_{z})$ of the mean spin vector $\langle 
\vec{S}\rangle $:%
\begin{eqnarray}
n_{x} &=\sin \vartheta \cos \varphi &=\left\langle \vartheta ,\varphi
\right\vert \op S_{x}\left\vert \vartheta ,\varphi \right\rangle /j  \nonumber \\
n_{y} &=\sin \vartheta \sin \varphi &=\left\langle \vartheta ,\varphi
\right\vert \op S_{y}\left\vert \vartheta ,\varphi \right\rangle /j  \label{n} \\
n_{z} &=\cos \vartheta &=\left\langle \vartheta ,\varphi \right\vert
\,\op S_{z}\left\vert \vartheta ,\varphi \right\rangle /j  \nonumber
\end{eqnarray}

A property of coherent states essential to us is the existence of
a special tangent plane orthogonal to the direction $\vec{n}$. If we define
a direction vector 
$\vec n_{\perp}(\chi)$ as $D(\vartheta,\varphi)T(\chi)\hat x$, we find $\vec n_{\perp}(\chi)\cdot \vec n=0$ for any $\chi$.

The observable 
\beq
\op S_\perp(\vartheta,\varphi;\chi)\equiv \vec n_\perp(\chi)\cdot \vec S=D(\vartheta,\varphi)\,T(\chi) \op S_x\,T^{-1}(\chi)\,D^{-1}(\vartheta,\varphi)
\eeq 
satisfies 
\beq
\Delta \left(\op S_\perp(\vartheta,\varphi;\chi)\right)^2=j  \label{inv_cs}
\eeq
independent of the angles $\vartheta,\varphi$ and $\chi$ when evaluated using $\ket{\vartheta,\varphi}$.

We will use the condition (\ref{inv_cs}) to fix the threshold of quantum
fluctuations and use this to define spin squeezing as was done by many
authors \cite{xxx} : a state of angular momentum $j$ is squeezed if there is
an orientation of $\vec{n}_{\perp }(\chi^*)$ in the tangent plane, defined for $T(\chi^*) \in {\cal H}$, for which 
\begin{equation}
\Delta \left(\op S_\perp(\vartheta,\varphi;\chi^*)\right)^2\leq \, j\,.
\end{equation}%
%
%
%

\subsection{SU(3) coherent states for $(\lambda,0)$ irreps.}

For $su(3)$ we consider symmetric irreducible representations of
the type $(\lambda ,0)$. The algebra is spanned by the six ladder operators $%
\hat{C}_{ij},i\neq j=1,2,3$ and two Cartan elements $\hat{h}_{1}=2\hat{C}%
_{11}-\hat{C}_{22}-\hat{C}_{33}$, $\hat{h}_{2}=\hat{C}_{22}-\hat{C}_{33}$. A
convenient realization is given in terms of harmonic oscillator creation and
destruction operators for mode $i$ by $\hat{C}_{ij}=\hat{a}_{i}^{\dagger }%
\hat{a}_{j}$ acting on the harmonic oscillator kets $|n_{1}n_{2}n_{3}\rangle 
$ with $n_{1}+n_{2}+n_{3}=\lambda $. One verifies, for instance, 
\beqa
[\op C_{ij},\op C_{k\ell}]&=&\op C_{i\ell}\delta_{jk}-\op C_{kj}\delta_{i\ell}\, ,\\
\hat{C}_{12}|n_{1}n_{2}n_{3}\rangle &=&\sqrt{(n_{1}+1)n_{2}}\, 
|n_{1}+1,n_{2}-1,n_{3}\rangle \,  .
\eeqa
$SU(3)$ elements are parametrized following a slight
adaptation of \cite{su3} by 
\begin{eqnarray}
R(\alpha _{1},\beta _{1},\alpha _{2},\beta _{2},\alpha _{3},\beta
_{3},\gamma _{1},\gamma _{2}) &=&R_{23}(\alpha _{1},\beta _{1},-\alpha
_{1})R_{12}(\alpha _{2},\beta _{2},-\alpha _{2})  \nonumber \\
&\times &R_{23}(\alpha _{3},\beta _{3},-\alpha _{3})\e^{-i\gamma _{1}\hat{h}%
_{1}}\,\e^{-i\gamma _{2}\hat{h}_{2}}
\end{eqnarray}%
where $R_{ij}(\eta ,\theta ,\varphi )$ is a transformation of the $SU(2)$
subgroup with subalgebra spanned by $\hat{C}_{ij},\hat{C}_{ji},\textstyle{%
\frac{1}{2}}[\hat{C}_{ij},\hat{C}_{ji}]$.

The highest weight state $|\lambda 00\rangle $ is invariant (up to a phase) under
transformations of the type $R_{23}(\alpha _{3},\beta _{3},-\alpha _{3})\e%
^{-i\gamma _{1}\hat{h}_{1}}\,\e^{-i\gamma _{2}\hat{h}_{2}}$, which generate
a $\mathcal{H}=U(2)$ subgroup. Coherent states are labeled by points on $%
SU(3)/U(2)\sim S^{4}$. Thus, using $\omega =(\alpha _{1},\beta _{1},\alpha
_{2},\beta _{2})$ as coordinates on $S^{4}$, we generate the coherent state $%
|\omega \rangle $ in the standard form \cite{perelomov, gilmore} as orbit of
the highest weight $|\lambda 00\rangle $ under the action of the
displacement operator on $S^{4}:$     
\begin{equation}
|\omega \rangle =D(\omega )|\lambda 00\rangle \equiv R_{23}(\alpha
_{1},\beta _{1},-\alpha _{1})R_{12}(\alpha _{2},\beta _{2},-\alpha
_{2})|\lambda 00\rangle \,.  \label{su3coherentstate}
\end{equation}%
This coherent state can also be represented as a product of $\lambda $
one-qutrit states 
\begin{eqnarray}
|\omega \rangle  &\propto &|\omega \rangle _{1}\otimes |\omega \rangle
_{2}\otimes \ldots \otimes |\omega \rangle _{\lambda }\,, \\
|\omega \rangle _{a} &=&\cos \textstyle{\frac{1}{2}}\beta _{2}|100\rangle
_{a}+\e^{i\alpha _{2}}\cos \textstyle{\frac{1}{2}}\beta _{1}\sin \textstyle{%
\frac{1}{2}}\beta _{2}|010\rangle _{a}  \nonumber \\
&&\qquad \qquad \quad +\e^{i(\alpha _{1}+\alpha _{2})}\sin \textstyle{\frac{1%
}{2}}\beta _{1}\sin \textstyle{\frac{1}{2}}\beta _{2}|001\rangle _{a}\,.
\end{eqnarray}

$|\omega \rangle $ is completely determined by a `` mean
vector'' $\vec{n}$ with (complex) components 
\begin{equation}
\vec{n}=\left( \langle \hat{C}_{23}\rangle ,\langle \hat{C}_{32}\rangle
,\langle \hat{C}_{12}\rangle ,\langle \hat{C}_{21}\rangle ,\langle \hat{C}%
_{13}\rangle ,\langle \hat{C}_{31}\rangle ,\langle \hat{h}_{1}\rangle
,\langle \hat{h}_{2}\rangle \right) \,.  \label{su3vector}
\end{equation}%
(A vector with $8$ real components is obtained using $\langle \hat{C}%
_{ij}\rangle +\langle \hat{C}_{ji}\rangle $ and $-i(\langle \hat{C}%
_{ij}\rangle -\langle \hat{C}_{ji}\rangle )$.) With 
\begin{equation}
T\equiv R_{23}(\alpha _{3},\beta _{3},-\alpha _{3})\e^{-i\gamma _{1}\hat{h}%
_{1}}\,\e^{-i\gamma _{2}\hat{h}_{2}}\in \mathcal{H}\,,  \label{Ttransfo}
\end{equation}%
it is easy to verify that the variance of the observable 
\begin{eqnarray}
\op{\cal K}(\alpha_3,\beta_3,\chi) &\equiv &T(\hat{C}_{13}+\hat{C}_{31})T^{-1} \\
&=&\left(\hat{C}_{13}+\hat{C}_{31}\right) \cos \textstyle{\frac{1}{2}}\beta _{3}\cos \frac{1}{2}\chi  \nonumber \\
&&-i\left(\hat{C}_{13}-\hat{C}_{31}\right)\cos \textstyle{\frac{1}{2}}\beta _{3}\sin \textstyle{\frac{1}{2}}\chi   \nonumber \\
&&-\ \left(\hat{C}_{12}+\hat{C}_{21}\right) \sin \textstyle{\frac{1}{2}}\beta _{3}\cos (\alpha _{3}-\textstyle{\frac{1}{2}}\chi ) \nonumber \\
&&-i\left(\hat{C}_{12}-\hat{C}_{21}\right)\sin \textstyle{\frac{1}{2}}\beta _{3}\sin (\alpha _{3}-\frac{1}{2}\chi
),  \label{frakK}
\end{eqnarray}%
where $\chi =6\gamma _{1}+\gamma _{2},$ when evaluated using the highest
weight state $|\lambda 00\rangle $, is $\lambda $ and independent of the
angles $(\alpha_3,\beta_3,\gamma_1,\gamma_2)$. Hence, the variance of 
\begin{equation}
\hat{{\cal K}}_\perp (\omega;\alpha_3,\beta_3,\chi) =D(\omega )\, \op{\cal K}(\alpha_3,\beta_3,\chi)\, D^{-1}(\omega )  \label{Gop}
\end{equation}%
when evaluated using the coherent state $D(\omega )|\lambda 00\rangle $, is
also independent of the ``direction'' $(\alpha_3,\beta_3,\chi)$ in the ``tangent hyperplane'' perpendicular to $\vec n$, and equal to $\lambda $.
Thus, we will use $(\Delta \hat{{\cal K}}_\perp (\omega;\alpha_3,\beta_3,\chi))^2=\lambda$ as our squeezing
threshold and define an $su(3)$ state $\vert \psi \rangle$ as squeezed if
there is an observable of the form $\hat{{\cal K}}_\perp (\omega;\alpha_3^*,\beta_3^*,\chi^*)$ for which 
\begin{equation}
(\Delta \hat{{\cal K}}_\perp (\omega;\alpha_3^*,\beta_3^*,\chi^*))^2<\lambda
\end{equation}
when evaluated in $\vert \psi \rangle$.

\section{Semiclassical squeezing}

Squeezing related to a given algebra of observables is understood to reflect
correlations (commonly called quantum correlations) between components of a
basis of an irrep. As mentioned before group transformations, 
obtained by exponentiating linear combinations of elements from the
algebra, produce rigid displacements of the basis states.  Correlations
between basis states cannot as a matter of definition be induced by such
group transformations. Rather, correlations can be either constructed through a
special preparation, or obtained as a result of non-linear (in terms of the
algebra of observables) transformations (usually from non-linear Hamiltonian
evolution) applied to initially uncorrelated systems.

In the case of large systems, it is convenient to analyze the evolution using
the phase-space approach. The reasons are twofold: we can not only represent
the initial state as a real-valued function and "draw" it (for some
appropriately chosen cuts) in the form a distribution ``covering'' some slices of the
phase-space, but more importantly also deduce many qualitative features of the time-evolution of
this distribution.  For a
wide class of quantum systems with a symmetry group ${\cal G}$, the phase-space
functions are defined through an invertible map \cite{brif}, so that we
associate to an operator $\hat{X}$ a phase-space symbol  
\begin{equation}
\hat{X}\mapsto W_{X}(\Omega )=\hbox{tr}(\hat{w}(\Omega )\hat{X}),
\label{map}
\end{equation}%
where the quantization kernel $\hat{w}(\Omega )$ is a Hermitian operator
defined on the classical manifold ${\cal G}/{\cal H}$ and
$\Omega $ denotes the phase-space coordinates. 

A  feature of this mapping is that the commutator of two
elements $\hat{X}$ and $\hat{Y}$ of the Lie algebra ${\mathfrak g}$  corresponding to the
group ${\cal G}$ is mapped to the Poisson brackets of the respective symbols:
\begin{equation}
\lbrack \hat{X},\hat{Y}]\propto \{W_{X},W_{Y}\}_P.
\end{equation}
The commutator of two generic operators is in general mapped to the
so-called Moyal bracket.  

For $SU(3)$ irreps of the type $(\lambda,0)$ and $\lambda
\gg 1$, and for sufficiently localized initial states in a class
dubbed `` semiclassical'' \cite{ballantain}, \cite{jex}, the short time dynamics can be
well described by the Liouville--type equation for the evolution of the Wigner
function: 
\begin{equation}
\partial _{t}W_\rho(\Omega )=\varepsilon \{W_\rho(\Omega ),W_{H}(\Omega
)\}_{P}+O(\varepsilon ^{3}),\quad   \label{ee1}
\end{equation}
where $W_\rho(\Omega )$ is the Wigner function, \emph{i.e.} the symbol of
the density matrix $\op \rho$ of the system, $W_{H}(\Omega )$ is the symbol of the Hamiltonian, and $%
\varepsilon $ is the so--called semiclassical parameter.  The Poisson bracket is in fact, the leading term in an expansion of the Moyal bracket in inverse powers
of the square root of eigenvalue of one of the Casimir operators in the $SU(3)$ irrep $(\lambda,0)$; 
we found that, for the mapping defined in \cite{GA} on $SU(3)/U(2)$ the semiclassical parameter
$\varepsilon$ is
\begin{equation}
\varepsilon =\frac{1}{2\sqrt{\lambda (\lambda +3)}}\, .
\end{equation}

The solution of (\ref{ee1}) can be written in general form as 
\begin{equation}
W_\rho(\Omega |t)=W_\rho(\Omega \left( t\right) ),  \label{semsol}
\end{equation}%
where $\Omega (t)$ denotes classical trajectories on $SU(3)/U(2)$. 
The approximation of dropping in Eqn.(\ref{ee1}) higher order terms in $\varepsilon$ describes well the initial stage
of the nonlinear dynamics, when self-interference is negligible. In physical applications, semiclassical
states often have the form of localized states (\textit{e.g.} coherent
states) and their "classicality" depends on non-invariance under the
transformations induced by symmetry subgroups of the (nonlinear) Hamiltonian
("classicality" is a subtle and delicate question not addressed
here) \cite{dodonov}, \cite{belov}.

The method of the Wigner functions allows us to calculate average values of the
observables giving drastically better results than the ``naive'' solution of
the Heisenberg equations of motion with decoupled correlators. On the other
hands, the quantum phenomena which are due to self-interference (like
Schr\"{o}dinger cats) are beyond the scope of this semiclassical
approximation.

\subsection{Phase space considerations}

From the parametrization of the coherent state of Eqn.(\ref{su3coherentstate}), we deduce a Poisson bracket on $S^4$, given by 
\begin{eqnarray}
\{f,g\} &=&\frac{4}{\sin \beta_{1}\sin^{2}\textstyle{\frac{1}{2}}\beta _{2}}
\left( \frac{\partial f}{\partial \alpha _{1}}\frac{\partial g}{\partial
\beta _{1}}-\frac{\partial g}{\partial \alpha _{1}}\frac{\partial f}{%
\partial \beta _{1}}\right)  \nonumber \\
&& -\frac{2\tan \textstyle{\frac{1}{2}}\beta _{1}}{\sin ^{2}\textstyle{\frac{%
1}{2}}\beta _{2}} \left( \frac{\partial f}{\partial \alpha _{2}}\frac{%
\partial g}{\partial \beta _{1}}-\frac{\partial g}{\partial \alpha _{2}}%
\frac{\partial f}{\partial \beta _{1}}\right)  \nonumber \\
&&+\frac{4}{\sin \beta _{2}} \left( \frac{\partial f}{\partial \alpha _{2}}%
\frac{\partial g}{\partial \beta _{2}}-\frac{\partial g}{\partial \alpha _{2}%
}\frac{\partial f}{\partial \beta _{2}}\right) ,
\end{eqnarray}
where $f$ and $g$ are any two functions on $SU(3)/U(2)$.

Following the prescription of \cite{GA}, we associate to an operator $\hat{X}
$ a phase-space symbol $W_{X}(\Omega )$ according to Eq.(\ref{map}). 
This map is linear on $\hat{X}$ so we only need to consider the phase
space symbols of a basis set constructed from $su(3)$ tensors $T_{(\nu
_{1}\nu _{2}\nu _{3})I}^{(\sigma ,\sigma )}$, which transforms under
conjugation by $g\in \mathcal{G}$ as the state $|(\sigma ,\sigma )\nu
_{1}\nu _{2}\nu _{3}I\rangle $ in irrep $(\sigma ,\sigma )$ transforms under 
$g$   Notational details
can be found in \cite{GA}.

$T_{(\nu _{1}\nu _{2}\nu _{3})I}^{(\sigma,\sigma )}$ takes the general form 
\begin{equation}
T_{(\nu _{1}\nu _{2}\nu _{3})I}^{(\sigma ,\sigma
)}=\sum_{n_{1}n_{2}n_{3}m_{1}m_{2}m_{3}}|n_{1}n_{2}n_{3}\rangle \langle
m_{1}m_{2}m_{3}|\tilde{C}_{n_{1}n_{2}n_{3};m_{1}m_{2}m_{3}}^{(\sigma \sigma
)(\nu _{1}\nu _{2}\nu _{3})I}
\end{equation}%
with $|n_{1}n_{2}n_{3}\rangle $ a state in the irrep $(\lambda ,0)$, $%
\langle m_{1}m_{2}m_{3}|$ an element in the dual representation $(0,\lambda
) $ and $\tilde{C}_{n_{1}n_{2}n_{3};m_{1}m_{2}m_{3}}^{(\sigma \sigma )(\nu
_{1}\nu _{2}\nu _{3})I}$ a coefficient closely related to the $su(3)$
Clebsch-Gordan coefficient occurring in the decomposition of elements in $%
(\lambda ,0)\otimes (0,\lambda )\rightarrow (\sigma ,\sigma )$. Note that
weights in $(\lambda,0)$ and $(0,\lambda)$ are multiplicity--free so the
triples $(n_1n_2n_3)$ and $(m_1m_2m_3)$ are enough to uniquely identify the
states. 

For irreps of the type $(\sigma,\sigma)$, some weights occur
multiple times and the label $I$, which specifies transformation properties
of the states under $SU(2)$ transformations generated by $R_{23}$, is
required to fully distinguish states with the same weights. 
The tensors  $T_{(\nu _{1}\nu _{2}\nu _{3})I}^{(1,1)}$ are
proportional to the generators of the $su(3)$ algebra. 

A generic tensor $T_{(\nu _{1}\nu _{2}\nu _{3})I}^{(\sigma ,\sigma )}$ is mapped to the phase
space function $T_{(\nu _{1}\nu _{2}\nu _{3})I}^{(\sigma ,\sigma )}\mapsto
W_{T_{(\nu _{1}\nu _{2}\nu _{3})I}^{(\sigma ,\sigma )}}(\Omega )$ 
\begin{equation}
W_{T_{(\nu _{1}\nu _{2}\nu _{3})I}^{(\sigma ,\sigma )}}(\Omega )=\sqrt{\frac{%
2(\sigma +1)^{3}}{(\lambda +1)(\lambda +2)}}D_{(\nu _{1}\nu _{2}\nu
_{3})I;(\sigma \sigma \sigma )0}^{(\sigma ,\sigma )}(\Omega )\,,
\end{equation}%
where $D$ is an $SU(3)$ group function defined in the usual way as the
overlap 
\begin{equation}
D_{(\nu _{1}\nu _{2}\nu _{3})I;(\sigma \sigma \sigma )0}^{(\sigma ,\sigma
)}(\Omega )=\langle ((\sigma,\sigma)\nu _{1}\nu _{2}\nu _{3}I\vert R(\Omega
)|(\sigma \sigma)\sigma\sigma\sigma 0\rangle
\end{equation}%
of two $su(3)$ states in the irrep $(\sigma ,\sigma )$.

The Wigner function corresponding to $\ket{\lambda 00}\bra{\lambda 00}$ is given by
\begin{eqnarray}
W_{\ket{\lambda 00}\bra{\lambda 00}}(\Omega ) &=&\sum_{\sigma =0}^{\lambda }\tilde{C}%
_{n_{1}n_{2}n_{3};n_{1}n_{2}n_{3}}^{(\sigma \sigma )(\sigma \sigma \sigma )0}%
\sqrt{\frac{2(\sigma +1)}{(\lambda +1)(\lambda +2)}}  \nonumber \\
&&\times \left( \frac{P_{\sigma +1}(\cos \beta _{2})-P_{\sigma }(\cos \beta
_{2})}{\cos \beta _{2}-1}\right) \,, \\
&\equiv &W_{\ket{\lambda 00}\bra{\lambda 00}}(\beta _{2})\,. \label{Whw}
\end{eqnarray}%
with $P_{\ell }$ a Legendre polynomial of order $\ell $. For $\lambda \gg 1$, we have found, with much similarity to the $SU(2)$ case \cite{wf limit}, 
that $W_{\ket{\lambda 00}\bra{\lambda 00}}(\beta _{2})$ is well approximated by 
\begin{equation}
W_{\ket{\lambda 00}\bra{\lambda 00}}(\beta _{2})\approx A\e^{\lambda (\cos \beta _{2}-1)},
\label{approxWF}
\end{equation}%
where $A=\frac{4\lambda ^{2}}{(\lambda +1)(\lambda +2)}$ is a constant
obtained so that the normalization condition 
\begin{eqnarray}
1 &=&\frac{(\lambda +1)(\lambda +2)}{8\pi ^{2}}\int d\Omega \,W_{\rho
_{\omega }}(\Omega )\,, \\
\int d\Omega &=&\int_{0}^{2\pi }d\alpha _{2}\int_{0}^{2\pi }d\alpha
_{1}\int_{0}^{\pi }\sin \beta _{1}d\beta _{1}\int_{0}^{\pi }\frac{1-\cos
\beta _{2}}{4}\sin \beta _{2}\,,
\end{eqnarray}%
is satisfied. The approximate expression (\ref{approxWF}) does not describes
very well the tail of the Wigner function, but for our purposes this is not essential.

For the coherent state $\ket{\omega}=R(\omega )|\lambda 00\rangle$, the density operator $\hat{\rho}_{\omega }=|\omega \rangle \langle \omega|$ is mapped to the Wigner function $W_{\rho_{\omega }}(\Omega )$ 
\begin{equation}
W_{\rho _{\omega }}(\Omega )=W_{\ket{\lambda 00}\bra{\lambda 00}}(\omega^{-1} \Omega)\, .
\end{equation}%

\subsection{Semiclassical evolution}

A simple Hamiltonian that leads to squeezing is 
\begin{equation}
\hat{H}=\hat{h}_{1}^{2}-\left( \frac{2\lambda +3}{5}\right) \hat{h}%
_{1}\,,\qquad \hat{h}_{1}=2\hat{C}_{11}-\hat{C}_{22}-\hat{C}_{33}\,,
\label{hamiltonian}
\end{equation}%
where the factor $\frac{2\lambda +3}{5}$ is chosen so that no
terms in $T_{(\nu _{1}\nu _{2}\nu _{3})I}^{(1,1)}$ appear in the expansion of $H$; this guarantees that no rigid motion
on the $S^{4}$ sphere is produced. This choice of $H$ is motivated on the
following physical grounds. The operator $\hat{h}_{1}$ is invariant under
the \emph{same} $U(2)$ transformations that leave the highest weight
invariant. Squeezing resulting from its evolution is thus a pure $SU(3)$
effect, distinct from $SU(2)$ correlations that are present in the
individual $U(2)$ subspaces contained in $(\lambda,0)$. Pure $SU(2)$ correlations
generated by non-linear Hamiltonians have been analyzed elsewhere \cite{su2ss}.
The symbol for this Hamiltonian is (up to a
constant factor) 
\begin{equation}
W_{H}=\frac{9}{40}\sqrt{(\lambda -1)\lambda (\lambda +3)(\lambda +4)}\left(
3+4\cos \beta _{2}+5\cos (2\beta _{2})\right) \,,  \label{W_H}
\end{equation}

We choose as initial state a coherent state with coordinates $\omega=(A_{1},B_{1},A_{2},B_{2})$ so it ``sits''
above the minimum of $H$ in (\ref{W_H}), \emph{i.e.} is located at 
$A_{1}=B_{1}=A_{2}=0$ and $B_{2}=\arccos (-1/5)$.
If we write the coset representative of $\omega ^{-1}\Omega $ as 
$(\bar{\alpha}_{1},\bar{\beta}_{1},\bar{\alpha}_{2},\bar{\beta}_{2})$, we
find for the coherent state and its symbol respectively:
\begin{eqnarray}
|\omega \rangle &=&R_{12}(0,B_{2},0)|\lambda 00\rangle \,,
\label{initialstate} \\
W_{\rho _{\omega }}(\Omega ) &=&W_{\ket{\lambda 00}\bra{\lambda 00}}(\bar{\beta}_{2})\, ,
\label{exactWF} \\
\cos \bar{\beta}_{2} &=&-1+2\cos ^{2}(\textstyle{\frac{1}{2}}B_{2})\cos ^{2}(%
\textstyle{\frac{1}{2}}\beta _{2})  \nonumber \\
&&+2\cos ^{2}(\textstyle{\frac{1}{2}}\beta _{1})\sin ^{2}(\textstyle{\frac{1%
}{2}}B_{2})\sin ^{2}(\textstyle{\frac{1}{2}}\beta _{2})  \nonumber \\
&&+\cos (\alpha _{2})\cos (\textstyle{\frac{1}{2}}\beta _{1})\sin (\beta
_{2})\sin (B_{2})\, , \label{cosbetabar}
\end{eqnarray}
with $W_{\ket{\lambda 00}\bra{\lambda 00}}$ given in Eqn.(\ref{Whw}).

Typical squeezing times scale as $t\sim \lambda ^{-p},$ $p>0$ and are much
shorter than self-interference times.  Hence, using $\ket{\omega}$ as initial state, we can use Eqn.(\ref{ee1}) to obtain
the approximate evolution as
\begin{equation}
\frac{dW_{\rho _{\omega }}}{dt}=\frac{9}{5}\sqrt{(\lambda -1)(\lambda +4)}%
(1+5\cos \beta _{2})\,\frac{\partial W_{\rho _{\omega }}}{\partial \alpha
_{2}}.
\end{equation}%
This in turn implies that the angle $\alpha _{2}$ evolves in time according
to 
\begin{equation}
\alpha _{2}(t)=\alpha _{2}+\frac{9}{5}\sqrt{(\lambda -1)(\lambda +4)}%
(1+5\cos \beta _{2})t\,,  \label{alpha2oft}
\end{equation}%
all other angles having no time dependence.
Thus, the time evolution of the system is obtained by the replacement
$\alpha _{2}\rightarrow \alpha _{2}(t)$ of Eqn.(\ref{alpha2oft}) in the argument $\cos \bar{\beta}_{2} $ of Eqn.(\ref{cosbetabar}) 
in the Wigner function of Eqn.(\ref{exactWF}):
\begin{equation}
W_{\rho _{\omega }}(\bar{\beta}_{2}|t)=W_{\rho _{\omega }}(\bar{\beta}%
_{2}(t))\, . \label{Wt}
\end{equation}%

\subsection{Semiclassical squeezing}

\begin{figure}[h!]
\begin{center}
\includegraphics[scale=0.70]{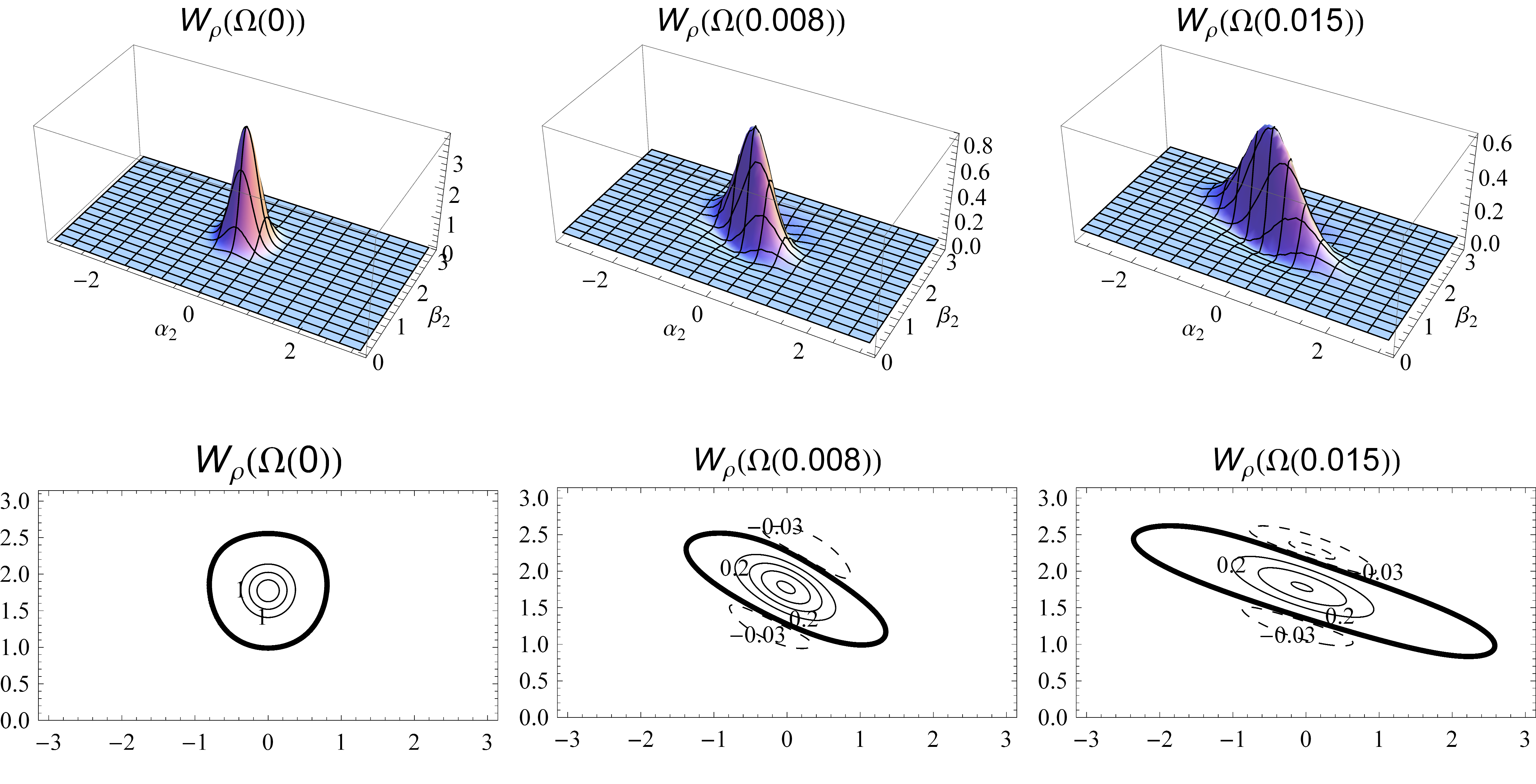}
\end{center}
\caption{Slices of the Wigner function for the initial state (\ref{initialstate}), evolved using the exact
evolution equation, for $t=0,0.008$ and $0.015$. The slices
are taken at $\alpha_1=\beta_1=0$. Note the small negative regions near the central peak at $t>0$.}
\label{exactgraph}
\end{figure}

\begin{figure}[h!]
\begin{center}
\includegraphics[scale=0.65]{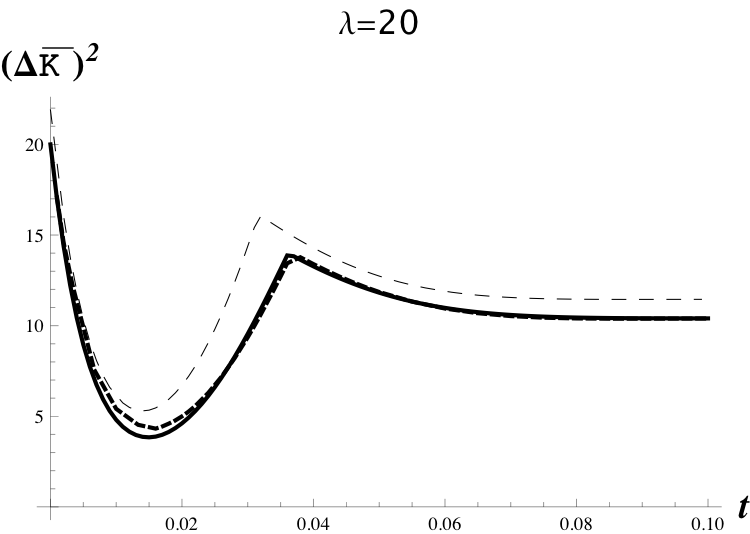}
\end{center}
\caption{The time evolution of the \emph{smallest} fluctuations of a system having as initial state the coherent
state (\ref{initialstate}). The full line is the smallest fluctuation of 
$\hat {\cal K}_\perp(\Omega;\alpha_3,\beta_3,\chi)$  calculated using the quantum evolution of (\ref{initialstate}), the thick dashed line was obtained using the classical
evolution of the exact Wigner function (\protect\ref{exactWF}) for (\ref{initialstate}), and the thin dashed line was obtained using the
classical evolution of the approximate Wigner function (\ref{approxWF}) for (\ref{initialstate}).}
\label{squeezinggraph}
\end{figure}

On Fig.\ref{exactgraph} we present as a 3D plot and as a contour plot the Wigner function for the
initial state (\ref{initialstate}), time--evolved using the exact quantum mechanical evolution
equation. The slices are taken at $\alpha _{1}=\beta_1=0$ and at specific values of $t=0, 0.008$ and $0.015$ as indicated.
(The value of $t=0.015$ is the time at which the fluctuation
of $( \Delta \hat {\cal K}_\perp(\Omega;\alpha_3^*,\beta_3^*,\chi^*)\left( t\right)) ^{2}$ reaches a minimum, as seen
on Fig.\ref{squeezinggraph}.)
One observes that initial coherent state is rapidly deformed
from its nearly Gaussian shape in $S^{4}$, spreads and leaves the tangent
hyperplane. In particular, small negative regions are generated in the
vicinity of the main peak.  

\begin{figure}[h!]
\begin{center}
\includegraphics[scale=0.70]{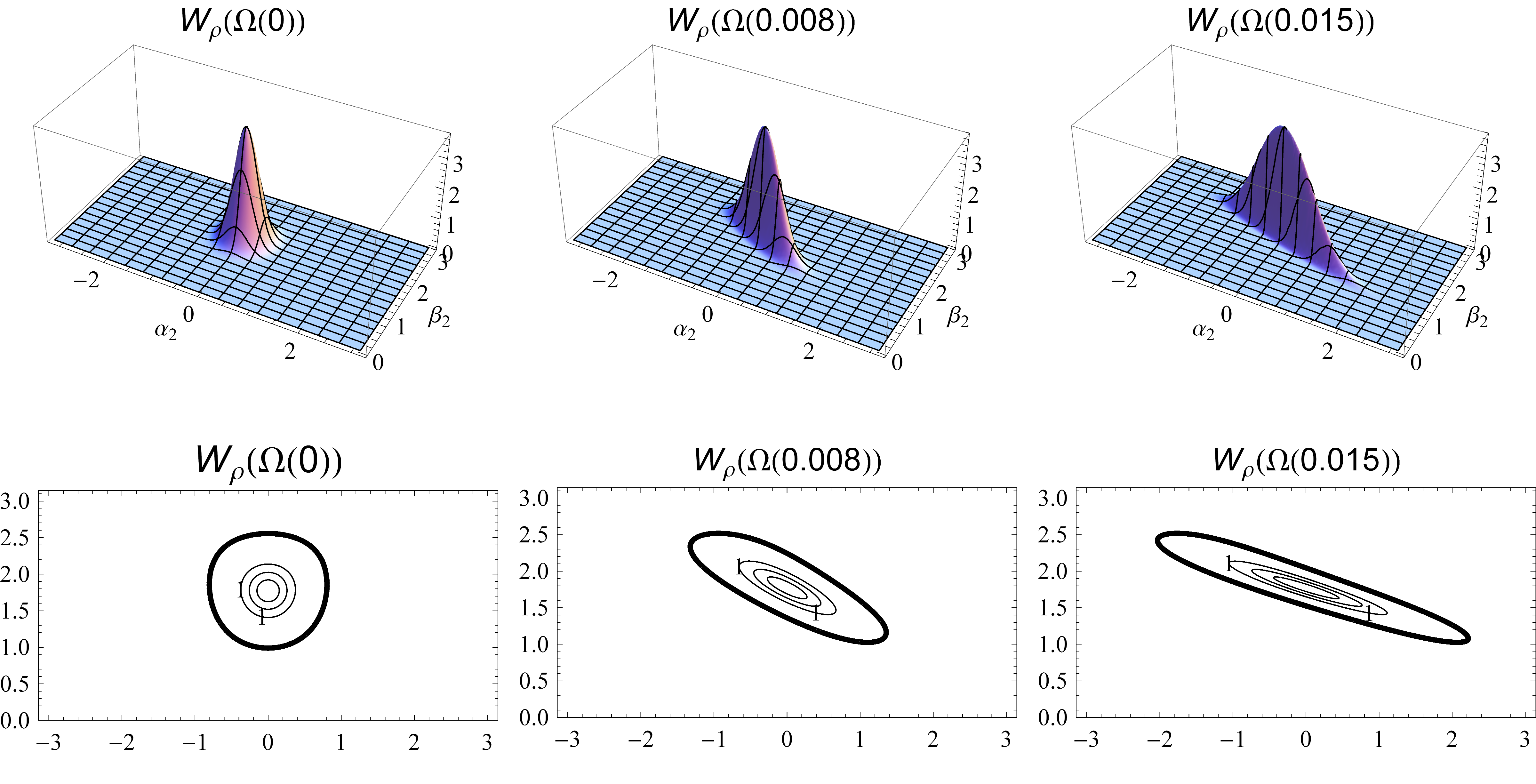}
\end{center}
\caption{Slices of the Wigner function for the initial state (\ref{initialstate}), evolved using the classical 
evolution equation, for $t=0,0.008$ and $0.015$. The slices
are taken at $\alpha_1=\beta_1=0$. There are no regions of where the function is negative.}
\label{semigraph}
\end{figure}

Fig.\ref{semigraph} illustrates the 3D and 
contour plots of slices of the Wigner function time-evolved using semiclassical evolution of the initial state.  
The times and slices are the same as for the exact evolution to facilitate comparisons.
Obviously we cannot observe negative regions in the Wigner function.

Fluctuations of the operator $\hat {\cal K}_\perp(\Omega;\alpha_3,\beta_3,\chi)$ of Eqn.(\ref{Gop}) are
invariant under $U(2)$ transformations $T$ when evaluated using the coherent
state $\vert \omega \rangle$ of Eqn.(\ref{initialstate}). If the quantum
correlations are induced by a non-linear Hamiltonian leaving stationary the
mean vector of Eqn.(\ref{su3vector}) characterizing $|\omega \rangle $, we
can use the same observables $\hat {\cal K}_\perp(\Omega;\alpha_3,\beta_3,\chi)$ to detect squeezing.
Operationally this means the fluctuations of $\hat {\cal K}_\perp(\Omega;\alpha_3,\beta_3,\chi)$ will now
depend on the parameters $\alpha _{3},\beta _{3},\chi =6\gamma _{1}+\gamma
_{2}$ of the transformation $T$ of Eqn.(\ref{Ttransfo}) through the
combinations of Eqn.(\ref{frakK}), in such a way that there may exist
"directions" parametrized by $\alpha _{3}^*,\beta _{3}^*,\chi^* $ in the tangent
hyperplane where the fluctuations are smaller than in the coherent state 
$|\omega \rangle $. It remains to select from those directions the one along
which the fluctuations are smallest to complete our definition of squeezing.

Average values and the fluctuations are computed using the standard
phase-space techniques, \emph{i.e.} integrating the symbols of 
$\hat {\cal K}_\perp(\Omega;\alpha_3,\beta_3,\chi)$ and its square with the time--evolved Wigner
function. Although the analytical integration can be done, the corresponding
expressions for $( \Delta \hat {\cal K}_\perp(\Omega;\alpha_3^*,\beta_3^*,\chi^*)\left( t\right)) ^{2}$ are formidable;  we will only provide numerical
results and compare in Fig. \ref{squeezinggraph} the results of exact
quantum mechanical calculations with those obtained from the Wigner function
method.

Figure \ref{squeezinggraph} shows the time--evolution of the \emph{smallest} fluctuations of $\hat {\cal K}_\perp(\Omega;\alpha_3,\beta_3,\chi)$ for the initial
coherent state (\ref{initialstate}) or its approximation (\ref{approxWF}) (where $\beta_2\to\bar\beta_2$) with $\lambda=20$ under the Hamiltonian 
$\hat{H}=\hat{h}_{1}^{2}-\frac{43}{5}\hat{h}_{1}$. The best squeezing direction $(\alpha_3^*,\beta_3^*,\chi^*)$
has been found though numerical optimization.

The results are typical, although the differences between the exact quantum
evolution and the classical evolutions decrease with $\lambda $. Through
numerical experiment, we have found that the location in time of the minimum
of $( \Delta \hat {\cal K}_\perp(\Omega;\alpha_3^*,\beta_3^*,\chi^*)\left( t\right)) ^{2}$ scales like $t_{min}\sim
\lambda^{-9/11} $ and the effective squeezing, defined as the ratio of the
minimum $( \Delta \hat {\cal K}_\perp(\Omega;\alpha_3^*,\beta_3^*,\chi^*)\left( t\right)) ^{2}/\lambda$, scales like $\lambda^{-1/3}$ for large values of $\lambda $.

\section{Conclusion}

We have shown that the reduction of fluctuations in the systems with $SU(3)$
symmetry can be achieved in a manner similar to the reduction in spin-like
systems: by correlating initially factorized coherent states via an
evolution generated by a Hamiltonian non-linear on the generators of the 
$su(3)$ algebra. 

We constructed the Hamiltonian in a such way that it does
not produce a rigid motion of the initial state, so we can use as observables
those having uniform fluctuations in a coherent state as a reference to
detect squeezing. Although we have not established a general criteria for 
$SU(3)$ squeezing, we have shown how quantum correlations (in the sense
described above) can lead to a reduction of fluctuations, which is reflected
through a specific deformation ("squeezing") of the initial coherent state.
It must be emphasized that, in quantum systems with higher symmetries,
different types of squeezing can be identified, and these types can be
conceptually different from the so-called one and two axis squeezing
typically found in spin-like systems. Here we used the Hamiltonian invariant
under $U(2)$ transformations and thus producing " true" , (\emph{i.e.} not
reducible to the $U(2)$-type interactions) $SU(3)$ correlations.

It should be also observed that in, contrast to spin-like systems, the exact
quantum mechanical calculations for physical models with $SU(3)$ symmetries
can be extremely cumbersome.  Thus, application of the phase-space methods
are extremely helpful not only for the geometrical interpretation and state
visualization, but also for estimating the evolution of systems in the limit
of large dimensions through the use of semiclassical calculations. In
particular, important physical effects such as squeezing, which originate
from non-trivial evolutions of collective qutrit fluctuations, can be
described in terms of semiclassical evolution of initial Wigner distribution
for suitable initial states. This is ultimately possible because the
approximate solutions (\ref{approxWF}) and (\ref{Wt}) describe well the
dynamics of initial semiclassical states for times of order $t\sim 1$, while
the major squeezing effect is achieved for \ times $t\sim \lambda ^{-p}$,
where $p>0$.

The work of ABK is partially supported by the Grant 106525 of CONACyT
(Mexico). The work of HDG is partially supported by NSERC of Canada. HTD
would like to acknowledge the financial support from Lakehead University.

\end{document}